\documentclass[twocolumn,showpacs,preprintnumbers,amsmath,amssymb,floatfix]{revtex4}
\usepackage{graphicx}% Include figure files
\usepackage{afterpage}% help with floats
\usepackage{amsmath}% for help with equations
\usepackage{dcolumn}% Align table columns on decimal point
\usepackage{bm}% bold math

\begin{document}

\preprint{APS/123-QED}

\title{$F$-resolved Magneto-optical Resonances at D1 Excitation of Cesium: \\ Experiment and Theory} 

\author{M.~Auzinsh}
\email{Marcis.Auzins@lu.lv}
\author{R.~Ferber}%
\author{F.~Gahbauer}%
\author{A.~Jarmola}%
\author{L.~Kalvans}%
\affiliation{
The University of Latvia, Laser Centre, Rainis Blvd., LV-1586 Riga, Latvia}%

%\affiliation{

\date{\today}% It is always \today, today,

\begin{abstract}
Bright and dark nonlinear magneto-optical resonances associated with the ground state Hanle effect have been studied experimentally and theoretically for D1 excitation of atomic cesium. This system offers the advantage that the separation between the different hyperfine levels exceeds the Doppler width, and, hence, transitions between individual levels can be studied separately. Experimental measurements for various laser power densities and transit relaxation times are compared with a model based on the optical Bloch equations, which averages over the Doppler contour of the absorption line and simultaneously takes into account all hyperfine levels, as well as mixing of magnetic sublevels in an external magnetic field. In contrast to previous studies, which could not resolve the hyperfine transitions because of Doppler broadening, in this study there is excellent agreement between experiment and theory regarding the sign (bright or dark), contrast, and width of the resonance. The results support the traditional theoretical interpretation, according to which these effects are related to the relative strengths of transition probabilities between different magnetic sublevels in a given hyperfine transition.

\end{abstract}
%32.60+i Zeeman and Stark effects
%32.80Xx Level crossing and optical pumping
%32.10Fn Fine and hyperfine structure
\maketitle

\section{\label{Intro:level1}Introduction}
\begin{figure}[htbp]
	\centering
		\resizebox{8cm}{!}{\includegraphics{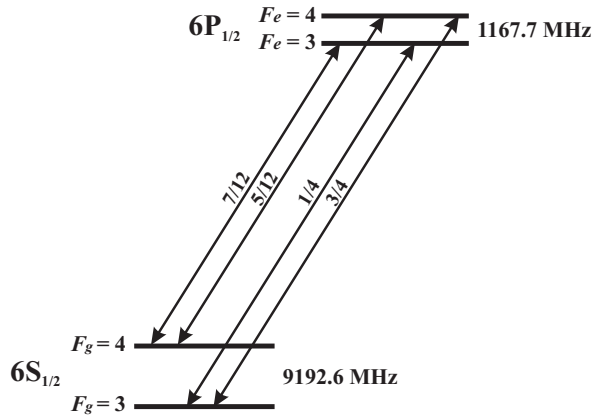}}
	\caption{\label{fig:levels} Transition of the Cesium D1 line. The fractions on the arrows indicate the relative transition strengths~\cite{Steck:cesium}.}
\end{figure}

	The nonlinear magneto-optical resonances associated with the ground state Hanle effect~\cite{Strumia} are a beautiful example of the manifestation ground state Zeeman coherences, and are also related to the effects of electromagnetically induced absorption and electromagnetically induced transparency~\cite{Budker:2004}. These sub-natural line width resonances appear in the fluorescence spectra of alkali atoms that are excited at the hyperfine transitions, and they may be "dark"~\cite{Alzetta:1976,Schmieder:1970} or "bright"~\cite{Dancheva:2000} Although straightforward theories have been proposed to explain the bright and dark resonances~\cite{Kazantsev:1984, Renzoni:2001, AlnisJPB:2001}, they have thus far eluded unambiguous verification because of experimental subtleties in the systems available for testing. We present an experimental and theoretical study of bright and dark resonances in the hyperfine transitions induced by D1 excitation of atomic cesium, 6$^2$S$_{1/2}$ ($F_g=3,4$) $\rightarrow$ 6$^2$P$_{1/2}$ ($F_e=3,4$), as shown in Figure~\ref{fig:levels}. Compared to previously studied systems, the cesium D1 line offers a much better defined test situation, because the separation between hyperfine levels of different total angular momentum $F$ exceeds the Doppler width. Therefore, the different transitions from ground state levels $F_g=3,4$ to excited state levels $F_e=3,4$ can be studied individually and the models can be verified in a more straightforward manner than heretofore possible. To our knowledge, the observations of the ground state Hanle effect in this system have thus far not been reported in the literature. 

	Sub-natural width resonances in alkali vapors due to the ground state Hanle effect were first reported by Schmieder and co-workers in 1970~\cite{Schmieder:1970}, although somewhat tentatively. The first unambiguous observation of ground state Hanle resonances is generally attributed to Alzetta and co-workers~\cite{Alzetta:1976}, who observed dark resonances in a beam of sodium atoms. As a result of their work, the effect came to be known as Coherent Population Trapping (CPT). Dark resonances are caused when atoms become trapped in a ground state sublevel of a particular magnetic quantum number $m$.

	Interest in coherent phenomena in atomic ground states intensified in the late 1990s because of applications to magnetometry~\cite{Scully:1992}, lasing without inversion~\cite{Scully:1989}, laser cooling~\cite{Aspect:1988}, electromagnetically induced transparency~\cite{Harris:1997}, and coherent information storage in halted light pulses~\cite{Fleischauer:2001,Liu:2001}. For many years only dark resonances had been observed, but in 2000 Dancheva and co-workers, working with alkali vapors in glass cells, reported bright resonances for the first time~\cite{Dancheva:2000}. This new type of resonance was quickly interpreted~\cite{Renzoni:2001,AlnisJPB:2001}. Unfortunately, subsequent experimental studies that applied these models could not always predict the correct sign of the resonance (bright or dark)~\cite{Alzetta:2001, Papoyan:2002}. One of the reasons for the difficulty was that Doppler broadening allowed several total angular momentum $F$ levels to participate in the transitions, while the numerical models were able to take into account only the cycling transition. Other experimental studies were plagued by difficulties in reproducing sufficiently homogeneous magnetic fields near zero~\cite{Alnis:2003}. Finally, a recent study~\cite{Andreeva:2007} of bright and dark resonances in cesium atoms confined in a nanometric cell was able to take advantage of the sub-Doppler properties of the nanometric cell~\cite{Sarkisyan:2001} to focus on transitions between individual $F$ levels. However, the nanometric cell adds additional subtleties that reverse the sign of some resonances, and the interpretation of these phenomena requires further study. Thus, although a beautiful and straightforward theoretical explanation of bright and dark ground state Hanle resonances has existed for some time, the overall situation has remained somewhat unsatisfactory because of ambiguous experimental results.

	We chose to work at the D1 line of cesium atoms confined in a glass cell. This system offers several advantages over the systems studied previously. For the cesium D1 line, the hyperfine splitting of the excited state is about 1.2~GHz (see Fig.~\ref{fig:levels}); therefore, the hyperfine transitions can be studied individually despite Doppler broadening. Unlike in the case of the thin cell, there is no need to account for additional effects, such as interactions with the walls. Therefore, it is much more straightforward to model the system and to compare the experimental and numerical results. Furthermore, we have continued to develop our modeling capability to take into account all $F$ levels that could be excited by tails of the laser radiation distribution, to average over all velocity groups, and to account for magnetic sublevel mixing in a magnetic field. We thus planned to apply a well-developed model to a simple system and to compare experimental results and theoretical expectations under a wide variety of experimental conditions. We could select the transition, vary the intensity of laser light and the transit relaxation rate (related to laser beam diameter), and observe the sign of the resonance, its contrast, and its width. The goal of our study was to gain confidence in the theoretical understanding of the bright and dark resonances and to try to settle outstanding doubts. 

\section{\label{Experiment:level1}Experimental Description}
\begin{figure}[htbp]
	\centering
		\resizebox{4cm}{!}{\includegraphics{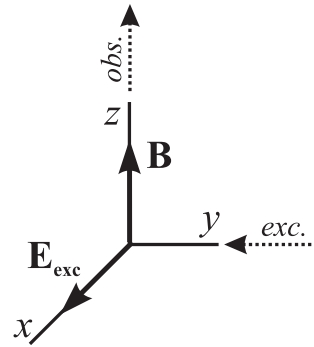}}
	\caption{\label{fig:geometry} Relative orienation of laser beam, laser polarization, magnetic field, and observation directions.}
\end{figure}

	Cesium vapor was confined at room temperature in a cylindrical glass cell with a diameter of approximately 5~cm. The cesium vapor was irradiated by linearly polarized laser radiation from an external cavity diode laser in the Littrow configuration, which was based on a Thorlabs L904P030 laser diode and was operated at the cesium D1 line (894.4 nm). The maximum laser power available at the entrance of the cell was 2.5 mW. This power could be attenuated incrementally up to 200 times by means of neutral density filters. Four beam profiles with cross-sectional areas of 0.125, 0.8, 2.0, and 4.7~mm$^2$ were obtained by means of lenses. The beam profile was characterized with a Thorlabs BP104-VIS beam profiler and the beam size was determined by considering as part of the beam all areas in which the power was greater than 50\% of the maximum power. Two lenses directed the laser-induced fluorescence onto a Thorlabs FDS-100 photodiode that was operated in photovoltaic mode. No polarizers were inserted between the cell and the photodiode, so that the fluorescence was observed regardless of polarization. The observation direction was perpendicular to the laser beam and to the direction of polarization of the laser (see Figure~\ref{fig:geometry}). The cell was located inside a three-axis Helmholtz coil system. The magnetic field in the direction of fluorescence observation was scanned by a Kepco BOP-50-8M bipolar power supply. The ambient magnetic field in the other two directions was compensated by the Helmholtz coils. 

	The magnetic field along the observation direction was slowly scanned in 200 or 400 discrete steps of duration 300~ms. At each step, five distinct fluorescence measurements were taken. Thus, a typical scan lasted either 60 seconds or 120 seconds. The laser beam was chopped at frequencies of several hundred Hertz for phase-sensitive detection. The fluorescence signal was amplified by a transimpedance amplifier and fed into an Ortholoc-SC 9505 two-phase lock-in amplifier. The time constant of the lock-in amplifier was 30~ms. The amplified signal was digitized by a National Instruments PCI-6024E data acquisition card and stored on a PC. The background was determined by moving the laser off-resonance, and this background was subtracted from the measured signals.

	The laser induced fluorescence was measured as a function of magnetic field for the transitions from the ground state hyperfine level with total angular momentum $F_g=3$ to the excited state hyperfine level with $F_e=3$, from $F_g=3$ to $F_e=4$, from $F_g=4$ to $F_e=3$, and from $F_g=4$ to $F_e=4$ (see Figure~\ref{fig:levels}). The transitions were identified with the help of a HighFinesse W/S~6 wavemeter, whose absolute accuracy was 600~MHz. No active stabilization was needed, since the laser tended to remain on a particular transition long enough to collect many scans. Results were averaged over several scans and are shown in Figures~\ref{fig:resonances} and \ref{fig:dark}, the background having been subtracted (see section~\ref{Results:level1}).

\section{\label{Theory:level1}Theory}
Bright and dark resonances can be explained qualitatively with a simple model that was proposed in~\cite{AlnisJPB:2001,Papoyan:2002,Renzoni:2001}. This model is based on the relative strengths of transition probabilities between different magnetic sublevels in a given hyperfine transition. A dark resonance signal is expected when the atoms are optically pumped into a non-absorbing coherent quantum state, which is destroyed by the applied magnetic field. This case applies to the transitions $F_g = 3 \rightarrow F_e = 3$, $F_g = 4 \rightarrow F_e = 3$, and $F_g = 4 \rightarrow F_e = 4$. The coherences created by the optical pumping are destroyed when the Larmor frequency of the ground state becomes comparable to the relaxation rate of the ground state, which in our experiment occurs roughly when the magnetic field is on the order of 0.1~G. More details on the qualitative model can be found in~\cite{AlnisJPB:2001,Papoyan:2002,Renzoni:2001}.

The actual theoretical model used in the calculations makes use of the quantum density matrix formalism. The Liouville equations (optical Bloch equations) for the density matrix were reduced to rate equations for magnetic sublevel populations and Zeeman coherences without losing the completeness of the description\cite{Blushs:2004}:
\begin{widetext}
\begin{multline}\label{rate1}
\dfrac{\partial\rho_{g_ig_j}}{\partial t} = \left(\Gamma_{p,g_ie_m} + \Gamma_{p,e_kg_j}^{\ast}\right)\underset{e_k, e_m}{ \sum }(d_1^{g_ie_k})^{\ast}d_1^{e_mg_j}\rho_{e_ke_m} - \\
-\underset{e_k,g_m}{%
\sum }\left[\Gamma_{p,e_kg_j}^{\ast} (d_1^{g_ie_k})^{\ast}d_1^{e_kg_m}\rho_{g_mg_j} +
\Gamma_{p,g_ie_k} (d_1^{g_me_k})^{\ast}d_1^{e_kg_j}\rho_{g_ig_m}\right] - \\ -i\omega_{g_ig_j}\rho{g_ig_j} + \underset{e_i,e_j}{\sum}\Gamma_{g_ig_j}^{e_ie_j}\rho_{e_ie_j} -
\gamma\rho_{g_ig_j} + \lambda\delta(g_i, g_j)
\end{multline}

\begin{multline}\label{rate2}
\dfrac{\partial\rho_{e_ie_j}}{\partial t} = \left(\Gamma_{p,e_ig_m}^{\ast} + \Gamma_{p,g_ke_j}\right) \underset{g_k, g_m}{\sum }d_1^{e_ig_k}(d_1^{g_me_j})^{\ast}\rho_{g_kg_m} - \\ - 
\underset{g_k,e_m}{
\sum }\left[\Gamma_{p,g_ke_j} d_1^{e_ig_k}(d_1^{g_ke_m})^{\ast}\rho_{e_me_j} +
\Gamma_{p,e_ig_k}^{\ast} d_1^{e_mg_k}(d_1^{g_ke_j})^{\ast}\rho_{e_ie_m}\right] 
- i\omega_{e_ie_j}\rho{e_ie_j} - \Gamma\rho_{e_ie_j},
\end{multline}
\end{widetext}
where $\rho_{g_ig_j}$ and $\rho_{e_ie_j}$ are the density matrix elements for the ground and excited states, respectively. The first term in (\ref{rate1}) describes the repopulation of the ground state and the creation of Zeeman coherences due to induced transitions, $\Gamma_{p,g_ie_j}$ and $\Gamma_{p,e_ig_j}^{\ast}$ represent the interaction strengths between the ground and excited states, and $d_1^{e_ig_j}$ is the dipole transition matrix element. The second term stands for the changes of ground state Zeeman sublevel population and creation of ground state Zeeman coherences due to light absorbtion. The third term describes the destruction of ground state Zeeman coherences by the external magnetic field. The fourth term describes the repopulation and transfer of excited state coherences to the ground state due to spontaneous transitions. The fifth and sixth terms show the relaxation and repopulation of the ground state due to non-optical reasons (mainly atoms flying in and out of the interaction zone, i.e., transit relaxation; thus, $\gamma$ is the inverse value of the time needed for atoms to fly through the laser beam and can be estimated from the thermal velocities of the atoms while $\lambda$ is the population supply rate to the particular magnetic sublevel due to transit relaxation).

In  equation~(\ref{rate2}) the first term stands for the light absorbing transitions from the ground to the excited state; the second term denotes induced transitions to the ground state; the third describes the destruction of ground state Zeeman coherences in the external magnetic field; and the fourth term denotes the rate of spontaneous transitions to the ground state.

The term $\Gamma_{p,g_ie_j}$ can be calculated as follows:

\begin{equation}
 \Gamma_{p,g_ie_j} = \frac{|\varepsilon_{\overline{\omega}}|^2}{\hbar^2}\frac{1}{\left[\left(\frac{\Gamma}{2} + \frac{\Delta\omega}{2}\right) \pm i\left(\overline{\omega} - \boldsymbol{k}_{\overline{\omega}}\boldsymbol{v} - \omega_{e_jg_i}\right)\right]}
 \label{gammaP1},
\end{equation}
where $\Gamma$ stands for the natural linewidth of the transition, $\Delta\omega$ is the linewidth of the exciting light, $\overline{\omega}$ is the frequency of the exciting light, $\boldsymbol{k}_{\overline{\omega}} \boldsymbol{v}$ is the energy shift due to the Doppler effect, and $\omega_{e_jg_i}$ is the actual energy difference between the particular ground and excited state magnetic sublevels, $\varepsilon_{\overline{\omega}}$ is the electric field strength of the laser radiation at the central laser frequency $\overline{\omega}$ and  $\frac{|\varepsilon_{\overline{\omega}}|^2}{\hbar^2}$ is proportional to the laser power density. Instead of the laser power density, the Rabi frequency appears in the numerical calculations; it is proportional to the square root of laser power density. The Rabi frequency depends on a range of parameters, such as as the reduced matrix elements and the correspondence of the laser line profile to the absorption profile, which were not known with sufficient accuracy. Therefore, to fit the theoretical results and observed signals, the conversion factor between the squared Rabi frequency and the laser power density was theoretically estimated~\cite{Alnis:2003} and afterwards fine tuned to find the best match between theory and experiment. Once the best conversion factor was found, the same is used for all the transitions.

The calculations have been made for the full system, which means that if the laser light is tuned, for example, to the hyperfine transition $F_g = 3 \rightarrow F_e = 3$, all other possible transitions ($F_g = 3 \rightarrow F_e = 4$, $F_g = 4 \rightarrow F_e = 3$, $F_g = 4 \rightarrow F_e = 4$) with their respective probabilities are also taken into account, as are the probabilities for the excited level to decay to both hyperfine levels of the ground state.

The dipole transition elements are calculated as $d_1^{e_ig_j} = \langle e_i|\boldsymbol{d}_1 \cdot\boldsymbol{e} |g_j\rangle$. Once the magnetic field is applied, all the hyperfine levels are split into magnetic Zeeman sublevels with different energies. Moreover, the initial quantum state $|F,m\rangle$ becomes a superposition of all possible hyperfine sublevels $F$ with the same magnetic quantum number $m$ denoted as $|\xi,m\rangle$. In general $g_i$ and $e_j$ states can be represented by

\begin{equation}
    |\xi, m\rangle= \underset{F}{\sum}c\left[\xi, F, m\right]|F,m\rangle,
    \label{mixing}
\end{equation}
where $c\left[\xi, F, m\right]$ are the coefficients of mixing. To obtain these coefficients and also the energy shifts of the nonlinear Zeeman effect, the energy matrix describing the interaction between hyperfine levels for each possible magnetic sublevel is constructed as follows:

\[ \left( \begin{array}{cccc}
\mu[F_0] & \mu[F_0, F_0 +1] & 0  & 0 \\
\mu[F_0, F_0 +1] & \mu[F_0 + 1 ] & \mu[F_0 + 1 , F_0 + 2] & 0\\
0 & \mu[F_0 + 1 , F_0 + 2] & \mu[F_0 + 2] & \ldots \\
0 & 0 & \ldots &  \ddots \\
\end{array} \right)\]

As can be seen, the interactions with $\Delta F = 1$ are taken into account; only these states are mixed by the magnetic dipole interaction. The elements of the Hamilton matrix can be calculated this way:
\begin{multline}
    \mu[F, m_F]=\frac{A_{0}}{2} C \\ + B_{0}\frac{3/4 C (C+1) - I
    (I+1)J(J+1)}{2I(2I-1)J(2J-1)}+ \mu_{B}g_F m_F B % \\
    \label{F0}
\end{multline}
\begin{multline}
    \mu[F, F - 1, m_F] = -\frac{\mu_B}{2}(g_J -  g_I)B \times \\ \times \sqrt{\left(\frac{(J+I+1)^2-F^2}{F}\right)\left(\frac{F^2-m_F^2}{F(2F+1)(2F-1)}\right)},
    \label{F1}
\end{multline}
where $A_0$ and $B_{0}$ are the magnetic dipole and electric quadrupole hyperfine constants, $C = F (F+1) - I (I+1) - J (J+1)$, $I$ is the spin of the atomic nucleus, $L$ is the total angular momentum of electrons, $\mu_B$ is the Bohr magneton, $g$ is the Land\'e factor, and $B$ denotes the external magnetic field strength. When the matrix is constructed, one must find its eigenvalues and eigenvectors, which represent the energy splittings and level mixing coefficients, respectively.

To make the model more precise, the results have been averaged over the Doppler profile in the following way: the width of the Doppler profile has been estimated, and the signal has been obtained by summing the results calculated at all possible Doppler energy shift values multiplied by appropriate statistical weights, which represent the number of atoms in a particular velocity group.

\section{\label{Results:level1}Results}

\begin{figure*}[htbp]
	\centering
		\resizebox{\textwidth}{!}{\includegraphics{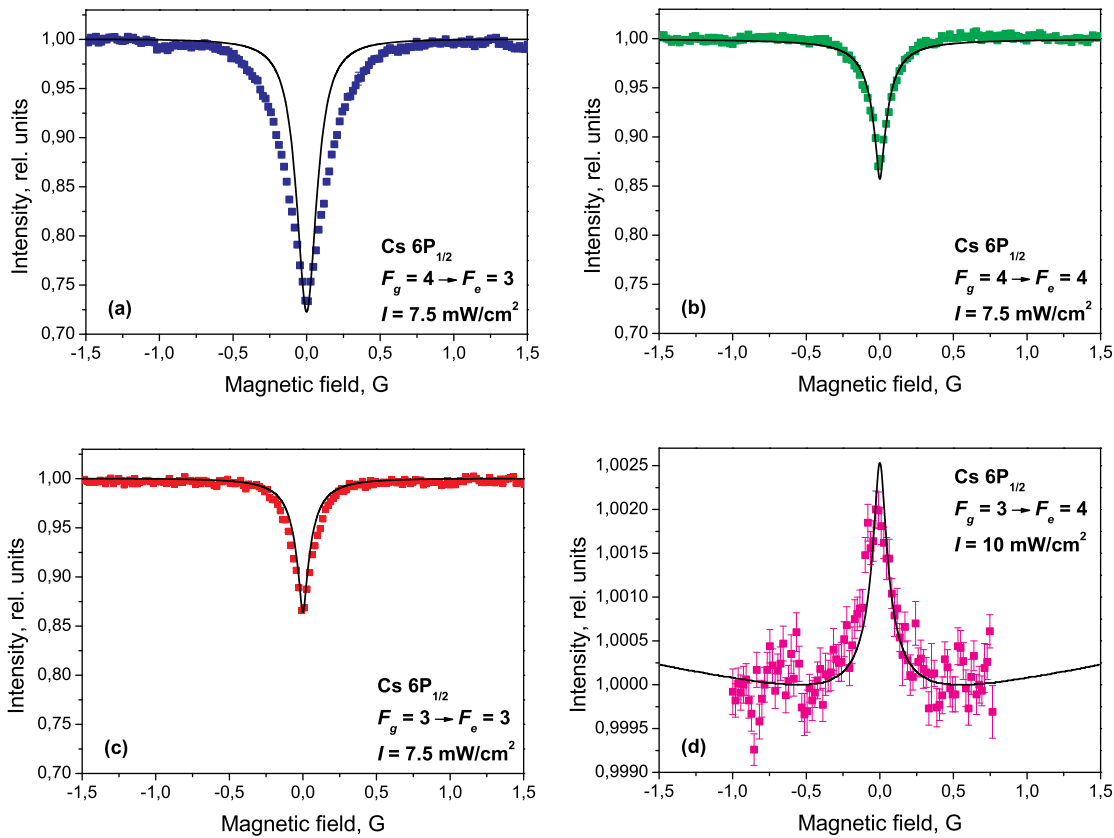}}
	\caption{\label{fig:resonances}Dark and bright resonances of the Cesium D1 transition. The intensity of the laser induced fluorescence in relative units is plotted against the magnetic field along the observation axis. Results are shown for the (a) $F_g=4 \rightarrow F_e=3$, (b) $F_g=4 \rightarrow F_e=4$, (c) $F_g=3 \rightarrow F_e=3$, and (d) $F_g=3 \rightarrow F_e=4$ transitions. Solid squares, experiment; solid line, theory. Note that the vertical axis is identical for (a), (b), and (c), but markedly differs in (d).}
\end{figure*}

Figure~\ref{fig:resonances} shows the intensity of laser induced fluorescence as a function of magnetic field along the observation direction for all transitions of the Cesium D1 line excited at comparable laser power densities. Dark resonances are observed at the $F_g=4 \rightarrow F_e=3$ (a), $F_g=4 \rightarrow F_e=4$ (b), and $F_g=3 \rightarrow F_e=3$ (c) transitions. At the $F_g=3 \rightarrow F_e=4$ (d) transition, a bright resonance is observed, as expected from the theoretical model. The contrast of the bright resonance is very small because the transition is not closed, but ``leaky'', which means that some excited atoms spontaneously decay to the other ground state and are lost. Therefore, they cannot participate in repeated absorption as would be required for the repopulation pumping cycle~\cite{Happer:1972}. The laser power density was approximately 7.5 mW/cm$^2$ for the dark resonances and 10 mW/cm$^2$ for the bright resonance. Although the laser power density was relatively high, agreement between theory and experiment is excellent, except for the case of the $F_g=4 \rightarrow F_e=3$ transition, in which case the theoretical model predicted a somewhat smaller width (see section~\ref{Analysis:level1} for a discussion). We point out that the contrast of the bright resonance is extremely small: approximately 0.2\%. Nevertheless, it is clearly visible and well-fitted by the theoretical model. 

\begin{figure*}[htbp]
	\centering
		\resizebox{\textwidth}{!}{\includegraphics{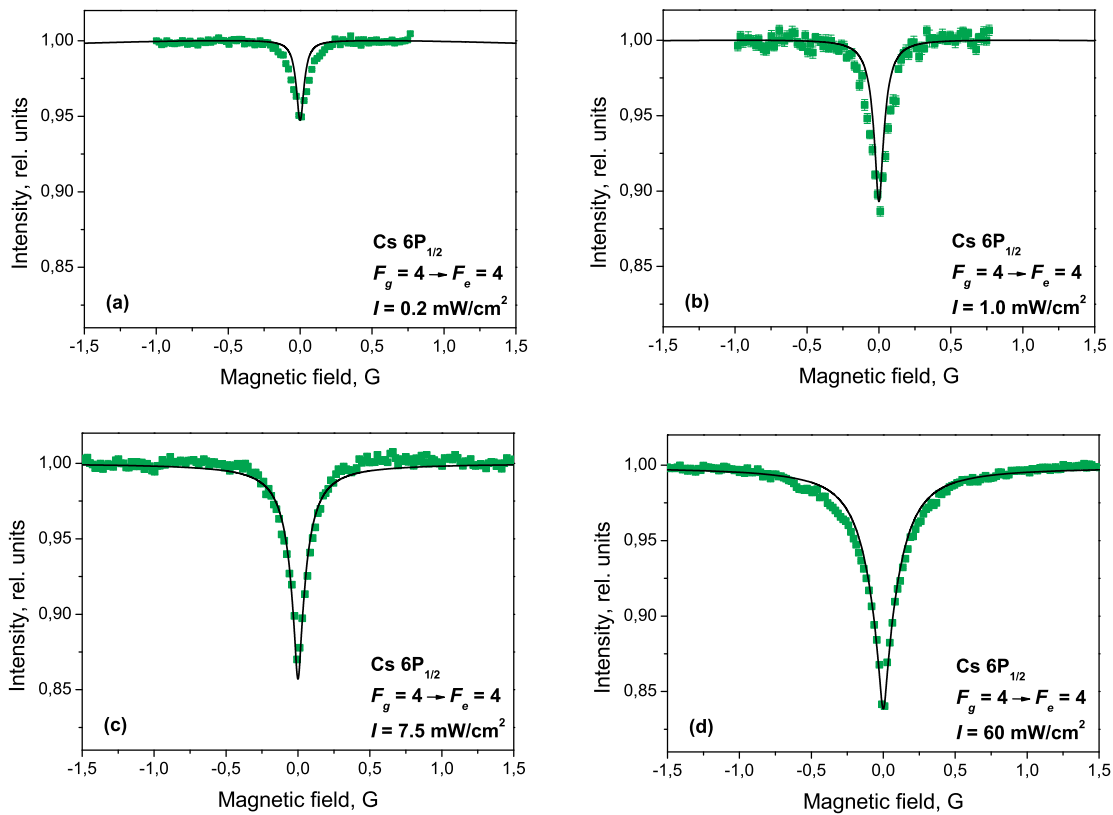}}
	\caption{\label{fig:dark} Dark resonance at the $F_g=4 \rightarrow F_e=4$ transition. The intensity of the laser induced fluorescence is plotted in relative units as a function of the magnetic field along the observation axis. Solid squares, experiment; solid line, theory.}
\end{figure*}

Figure~\ref{fig:dark} shows typical results for the case of a dark resonance. The intensity of the laser induced fluorescence is plotted as a function of the magnetic field along the observation direction. Experimentally measured values are represented by solid squares, while the black line shows the results of the theoretical calculation. In this case the $F_g=3 \rightarrow F_e=3$ transition has been studied for various power densities of laser radiation in the range from 0.1 mW/cm$^2$ to 120 mW/cm$^2$. The intensity scale has been arbitrarily normalized to unity for signal levels at fields far from the resonance, where the fluorescence intensity does not appear to depend on the magnetic field, at least to first order. The relationship between the magnetic field and the measured current in the coils was determined for high field values by means of a three axis Hall probe manufactured by Senis GmbH of Switzerland. The zero field point is assumed to be at the resonance position. Figure~\ref{fig:dark} shows that the contrast of the resonance increases as the laser intensity increases. By contrast we mean the ratio between the minimum fluorescence intensity at the resonance position and the fluorescence intensity far from the resonance. The phase sensitive detection eliminated any background that was not associated with the laser. We determined the background that was due to scattered laser light by measuring the signal when the laser was tuned far away from any transition. There was an additional background-like component due to scattered fluorescence light. Since the Hanle effect causes a spatial redistribution of fluorescence intensity, any stray fluorescence light that is detected will act as a kind of background and will tend to reduce the signal contrast. The magnitude of this effect was determined by fitting experimental curves to the theory to obtain a single background parameter for each transition; its magnitude was on the order of 10\% of the signal.

\section{\label{Analysis:level1}Analysis and Discussion}
\begin{figure*}[htbp]
	\centering
		\resizebox{\textwidth}{!}{\includegraphics{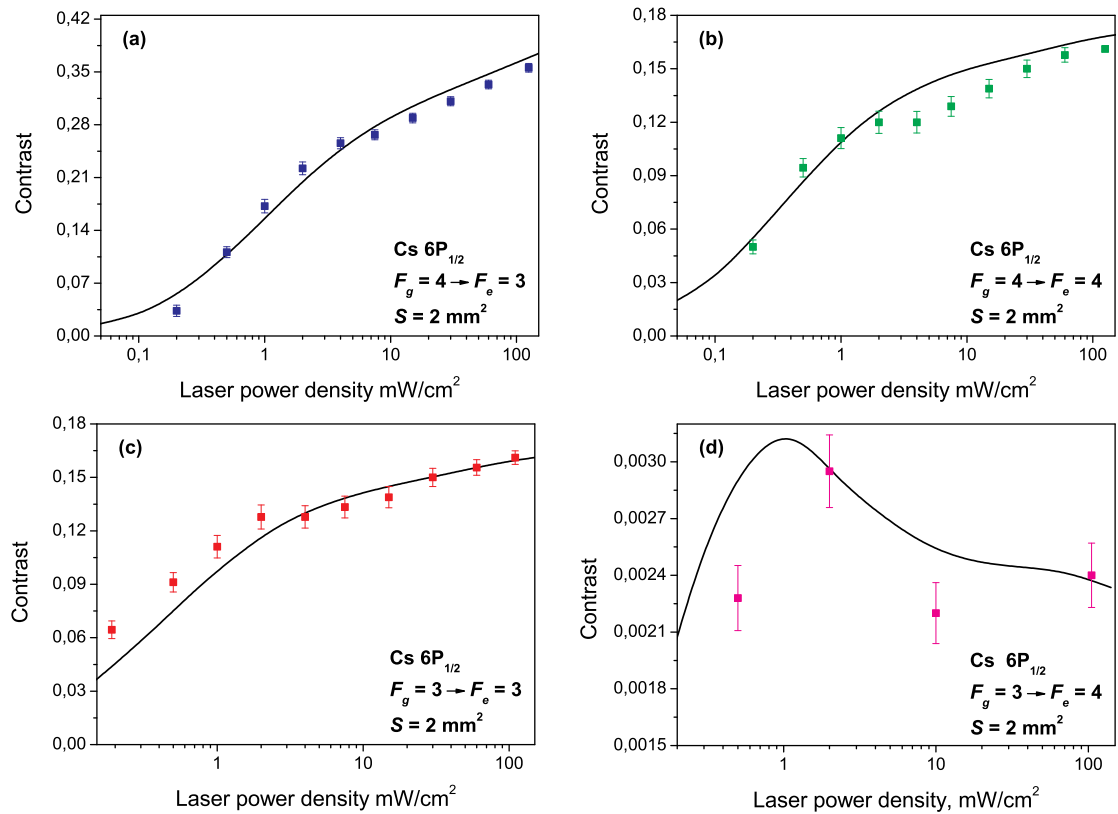}}
	\caption{\label{fig:contrast}Contrast of bright and dark resonances as a function of laser power density. The contrast of the resonances is plotted as a function of laser power density for the same cesium D1 transitions as in Fig.~\ref{fig:resonances}. Solid squares, experiment; solid line, theory. Note that the laser power density is plotted on a logarithmic scale. The $F_g=3 \rightarrow F_e=4$ transition (d) is a bright resonance, whereas the others are dark. The cross-sectional area of the beam $S$ was 2 mm$^2$.}
\end{figure*}

	The theoretical model was tested under a wide range of conditions in order to determine the accuracy of the model itself and to understand better the nature of bright and dark resonances. Figure~\ref{fig:contrast} shows the resonance contrast as a function of laser intensity for all of the transitions of the Cesium D1 line. The bright resonance at the $F_g=3 \rightarrow F_e=4$ transition (d) displays a behavior that differs markedly from that of the other transitions, which are dark. The contrast of the dark resonances increases monotonically as the laser power density increases, and approaches a saturation value. The reason for the increase in contrast with laser power density is that, as the power density increases, more and more atoms are pumped into a non-interacting magnetic sublevel of the original ground state. One would expect saturation, just as in any optical pumping phenomenon.
	
	The contrast of the bright resonance displays a maximum around a laser power density of 1 mW/cm$^2$ and then decreases. This behavior follows from the fact that, in the case of bright resonances, there are no non-interacting magnetic sublevels of the ground state. Instead, bright resonances are produced by changes in the ground state population distribution of atoms that are cycling between a ground and an excited state. If all excited atoms were to decay spontaneously to the original ground state level, the contrast of bright resonances would increase with increasing laser power density until some saturation value~\cite{Papoyan:2002}, just as in the case of the dark resonances. However, on the cesium D1 transition $F_g=3 \rightarrow F_e=4$, atoms can decay spontaneously to both ground state levels $F_g=3$ and $F_g=4$. This possibility is described as leakage. As a result of sufficiently strong laser power density, a substantial part of the atoms is ``pumped'' from the ground state level $F_g=3$ to the ground state level $F_g=4$. As a result, the atoms that are lost to the $F_g=4$ ground state level no longer contribute to the bright resonance, and the resonance contrast decreases. It is quite remarkable in a ``leaky'' system that bright resonances are observed at all. It is even more remarkable that there is such excellent agreement between theory and experiment for such a subtle effect. 

In the case of the dark resonance at the $F_g=4 \rightarrow F_e=3$ transition, the measured shape deviates slightly from the shape predicted by the model. We believe that the resason for this discrepancy is that optical pumping effects were stronger at this transition and, therefore, our treatment of the ground state relaxation rate was no longer fully adequate. In the $F_g=4 \rightarrow F_e=3$ transition there are two non-absorbing magnetic sublevels in the ground state, whereas in the case of the $F_g=4 \rightarrow F_e=4$ and $F_g=3 \rightarrow F_e=3$ transitions there is only one non-absorbing magnetic sublevel in the ground state. Also, the line strength of the $F_g=4 \rightarrow F_e=3$ transition is greater than is the case for the other two dark resonances (see Fig.~\ref{fig:levels}); therefore, pumping is more effective at the same intensity. Indeed, the contrast of the resonance at the $F_g=4 \rightarrow F_e=3$ transition is about twice as great as the contrast of the other two dark resonances. The width of the resonances is mainly determined by the ground state relaxation rate, which in this case is the transit relaxation rate. It was theoretically demonstrated that the dynamics of the transit relaxation from strong optical pumping deviate substantially from an exponential~\cite{Auzinsh:1983} and cannot be described by a single rate constant $\gamma$ as in our model (see equation~\ref{rate1}). If needed, the nonexponential transit relaxation rate could, in principle, have been included in the theoretical model, as was done peviously in~\cite{Auzinsh:1983}. However, available computational resources allowed us to include into the model either realistic transit relaxation dynamics or the Doppler averaging and magnetic sub-level mixing in an external magnetic field. The latter effects were estimated to be more important and were therefore incorporated in the model. We believe that this approach was fully justified by the results presented in this paper.	

\begin{figure*}[htbp]
	\centering
%		\resizebox{\textwidth}{!}{\includegraphics{Fig6}}
		\resizebox{8cm}{6cm}{\includegraphics{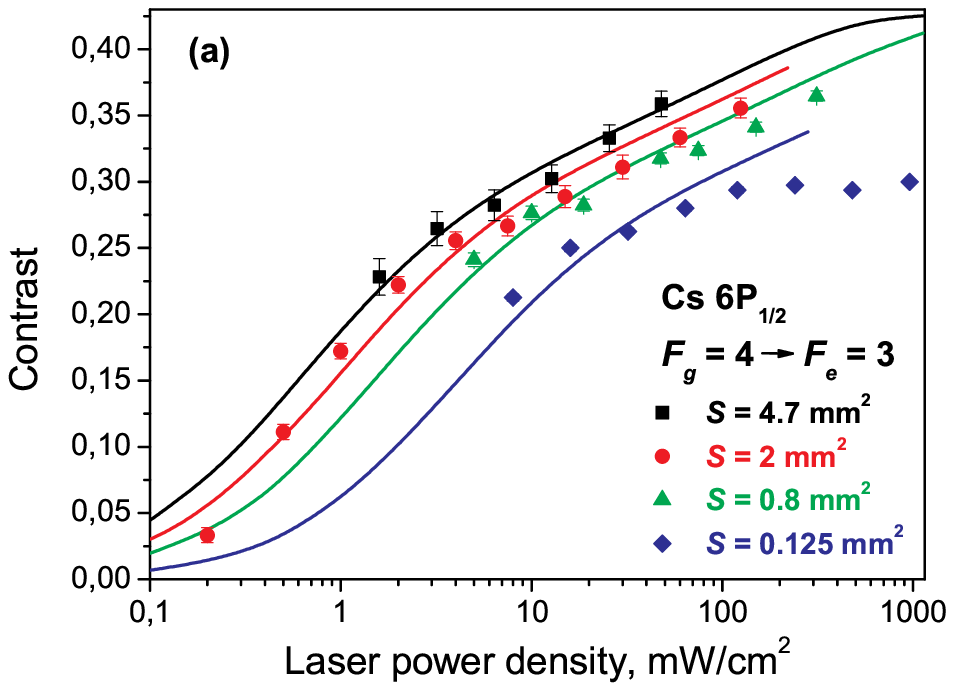}}
		\resizebox{8cm}{6cm}{\includegraphics{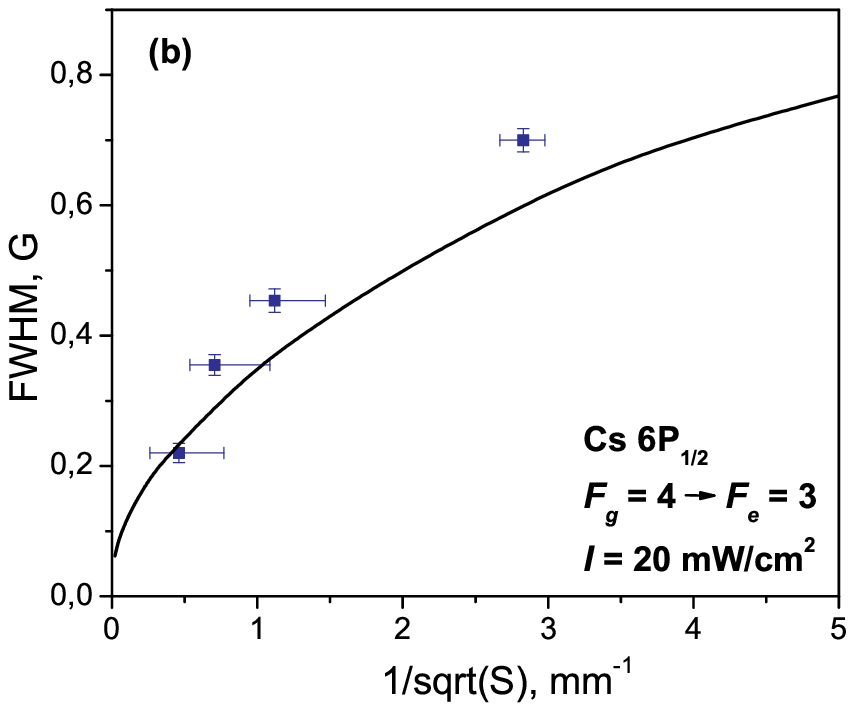}}
	\caption{\label{fig:beam}Resonance width and contrast as a function of beam properties.  In (a) the contrast is plotted as a function of laser power density for various beam cross sections. The points of different shapes correspond to the experiment. The lines correspond to calculations; the sequence of the lines from top to bottom corresponds to the sequence of the points in the legend. In (b) the width is plotted as a function of the beam dimension for the $F_g=4 \rightarrow F_e=3$ transition. The asymmetric beam was parametrized by a single dimension that was the square root of the cross-sectional area $S$. The horizontal error bars reflect the asymmetric nature of the beam. }
\end{figure*}

	In Fig.~\ref{fig:beam}(a) the resonance contrast is plotted against laser power density for various beam sizes. The theoretical curves agree quite well with the measured points for the larger beam sizes. Deviations become more and more significant as the cross-sectional area $S$ of the beam decreases, probably because the uncertainty in the transit relaxation time increases for small beam dimensions. At our experimental conditions, the ground state collisional relaxation rate can be estimated to be on the order of $\gamma_{col} \approx $ 100 s$^{-1}$, and thus it can be neglected. Considering the fact that the theoretical model was developed for a somewhat idealized beam profile, the agreement is quite satisfying.
	
	We also studied how the resonance width varies with the dimensions of the laser beam. The dimensions of the laser beam are related to the transit relaxation time. The width (FWHM) of the $F_g=3 \rightarrow F_e=4$ transition is plotted as a function of beam dimension in Fig.~\ref{fig:beam}(b). The intensity was held constant at 20~mW/cm$^2$. The points correspond to experimentally measured values. The line is the result of the theoretical model. The theoretical model uses a transit relaxation time determined from the beam dimensions and the thermal velocity of the atoms. The model assumes that the beam had a sharp cut-off. The real beam did not have a sharp cut-off, and its profile was generally asymmetric. We expect the smaller dimension to dominate the relaxation time. Nevertheless, since it was somewhat difficult to define the exact beam dimension in our experimental conditions, we plot error bars to indicate the maximum and minimum dimensions. The experimental results show good qualitative agreement with the theoretical model. As expected, the width tends to zero as the transit relaxation time tends to zero, and increases monotonically with increasing transit relaxation time. 

\section{\label{Conclusion:level1}Conclusion}
Bright and dark resonances were studied experimentally and theoretically for the first time on the cesium D1 transitions. This system was interesting because, unlike in other systems studied until now, each distinct $F$ level could be experimentally resolved and studied individually. Therefore, this transition can be used as a test system to understand better the formation of dark and bright resonances and to study their properties. Three dark and one bright resonance were observed. Never before had bright resonances been observed at transitions that were partially open (or leaky). The observed sign of the resonances (bright or dark), as well as their contrast, were exactly as expected according to our theoretical model, which was based on the optical Bloch equations. The model included magnetic sublevel mixing in the magnetic field as well as averaging over the Doppler profile and the influence of the other $F$ levels due to the Doppler effect and off-resonance absorption, even though it should to be small. Moreover, the theoretical description was extremely successful at reproducing not only the correct sign of the resonance, but also the width and contrast, even in cases were the contrast was on the order of a fraction of a percent. Only at high pumping rates were there some deviations in the widths of measured and calculated resonance signals, but even here the agreement in the contrast was very good. The slight deviations that did occur could be explained by the non-exponential character of the transit relaxation at high laser light intensity. We believe that these results establish that the assumptions on which the theoretical model was based are essentially correct. The model will therefore be useful in understanding more subtle effects in more complex systems. 
	
	In the event that ground state Hanle resonances in alkali vapors could serve as the basis for optical switches or adaptive optics (see, for example,~\cite{Yeh:1982}), it will be necessary to have a good grasp of the influence of all possible system parameters. This work has demonstrated that it is possible to construct a detailed and robust theoretical model that accurately models the ground state Hanle effect. Such a model would be indispensible in the design of optical devices.  

%\clearpage

\begin{acknowledgments}
We would like to thank Maris Tamanis for invaluable assistance with the experiments. We thank Janis Alnis for constructing the diode lasers and Robert Kalendarev for preparing the cesium cell. We thank Christina Andreeva and Aram Papoyan for useful discussions. We acknowledge support from the Latvian National Research Programme in Material Sciences Grant No.~1-23/50, the University of Latvia grant Y2-ZP04-100, the ERAF grant VPD1/ERAF/CFLA/05/APK/2.5.1./000035/018, and the INTAS projects 06-1000017-9001 and 06-1000024-9075. A.~J., F.~G., and L.~K. acknowledge support from the ESF project. 
\end{acknowledgments}
\bibliography{cesium}% Produces the bibliography via BibTeX.
\end{document}